\documentclass[a4paper,11pt]{article}
\usepackage{pos}

\title{Temporal and Spectral Analysis of 1ES 2344+514 in Two Flaring States Observed by VERITAS}
 \ShortTitle{VERITAS Analysis of 1ES 2344+514 During Two Flaring States}

\author*[a]{Connor Poggemann}
\author[a]{Jodi Christiansen}

\affiliation[a]{California Polytechnic State University,\\
  San Luis Obispo, CA, USA}

\onbehalf{for the VERITAS Collaboration} 

\emailAdd{cpoggema@calpoly.edu}
\emailAdd{jlchrist@calpoly.edu}

\abstract{VERITAS observed the bright blazar 1ES 2344+514 during two flaring periods, one from Dec. 17 to Dec. 18, 2015 (MJD 57373-57374) with a peak flux of $\sim$$60\%$ of the Crab and another from Nov. 28 to Dec. 3, 2021 (MJD 59546-59551) with a peak flux of $\sim$$20\%$ of the Crab. This blazar, located at a redshift of $z=0.044$, is classified as an extreme high-frequency-peaked BL Lacertae object (HBL). It is known to be variable, including several previous day-scale flares: Whipple on Dec. 20, 1995, VERITAS on Dec. 7, 2007, and MAGIC on Aug. 11, 2016. The VERITAS near-nightly monitoring of 1ES 2344+514 during the 2015-2016 and 2021-2022 seasons provides good coverage of the pre- and post-flare flux as well as the rise/fall time of the flares. We present the multiwavelength light curves of each flare as well as the very high-energy spectra in the two flare states and the two pre-flare states.}

\ConferenceLogo{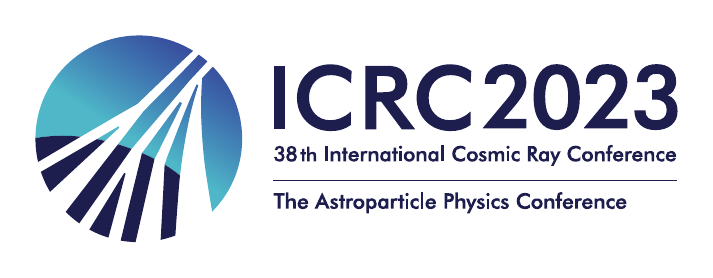}

\FullConference{%
38th International Cosmic Ray Conference (ICRC2023)\\
  26 July - 3 August, 2023\\
  Nagoya, Japan}

\begin{document}
\maketitle

\section{Introduction}

The blazar 1ES 2344+514 is highly variable and has been classified as an extreme high-frequency-peaked blazar (EHBL). It is located at a redshift of $z=0.044$ \cite{redshift} with coordinates RA: 23h 47' 04.836715" and Dec: +51d 42' 17.88123" (J2000) \cite{location}. The low redshift and the inverse-Compton peak at very-high energy (VHE) contribute to its observed brightness by imaging atmospheric Cherenkov telescopes (IACT). The Whipple collaboration published its discovery in 1998 \cite{whipple}. This was the 3rd blazar detected at energies above 300 GeV, after Markarian 421 and Markarian 501.

Flares from 1ES 2344+514 have been frequently detected at VHE. Whipple observed a flare on December 20, 1995 with an integral flux of (E > 350 GeV) 63\% of the Crab Nebula flux \cite{whipple}. VERITAS observed a flare on December 7, 2007 with an integral flux of (E > 300 GeV) 48\% of the Crab Nebula flux \cite{veritas2007}. MAGIC observed a flare on August 11, 2016 with an integral flux of (E > 300 GeV) 55\% of the Crab Nebula flux \cite{magic2016}. Due to its brightness and activity, 1ES 2344+514 is an ideal laboratory to observe high energy radiation and test multiwavelength models.

VERITAS observed 1ES 2344+514 regularly between 2007 and the present. The observations between 2007-2015 showed significant variability of yearly average fluxes \cite{allen}. In this proceeding we report two flaring periods, one from December 17 to 18, 2015 (57373-57374 MJD) with a peak flux of $\sim$$60\%$ of the Crab and another from November 28 to December 3, 2021 (59546-59551 MJD) with a peak flux of $\sim$$20\%$ of the Crab. Extensive monitoring of the source provides good coverage of the pre- and post-flare flux as well as the rise/fall time of the flares. We present the analysis of the light curves and spectra of each flare.

\section{VERITAS Observations}

The Very Energetic Radiation Imaging Telescope Array System (VERITAS) is a ground-based instrument at the Fred Lawrence Whipple Observatory (FLWO) in Arizona, USA. The array uses four 12-m imaging Cherenkov telescopes each mounted with a 499 photomultiplier tube camera. VERITAS is sensitive to VHE (100 GeV - 10 TeV) particle showers in the atmosphere. Its energy resolution is approximately $15\%$, and its angular resolution is approximately $0.08^{\circ}$ at 1 TeV \cite{holder, drc}.

VERITAS observed 1ES 2344+514 for a total of 11 hours during the 2015-16 season (7283-57375 MJD). Observations totaled 17 hours during the 2021-22 season (59486-59581 MJD). This data was taken under the VERITAS blazar monitoring campaign with the intent to record long-term quality baseline spectra as well as detect flares, should they occur. 

The air showers were reconstructed using the Image-Template Method \cite{drc} followed by a gamma-ray event selection. The selection criteria were optimized on the Crab Nebula scaled to 1\% of its nominal strength. Source event candidates were selected in a circular "on" region of radius $0.0707^{\circ}$. Average background events were determined from multiple reflected "off" regions chosen with the same offset from the center of the camera as the on-region \cite{wob}. The background events in the on-region were estimated using the ratio of the "on" and "off" angular areas, $\alpha$.
We found 635 "on" events and 1670 "off" events with an $\alpha$ of 0.06035, resulting in a peak significance of $35.0\sigma$ during the 2015-16 season. During the 2021-22 season, we found 342 "on" events and 2092 "off" events with an $\alpha$ of 0.05969, resulting in a peak significance of $15.7\sigma$.

\begin{figure}[b]
    \includegraphics[width=77mm,trim={4.5cm 10cm 5.5cm 10cm},clip]{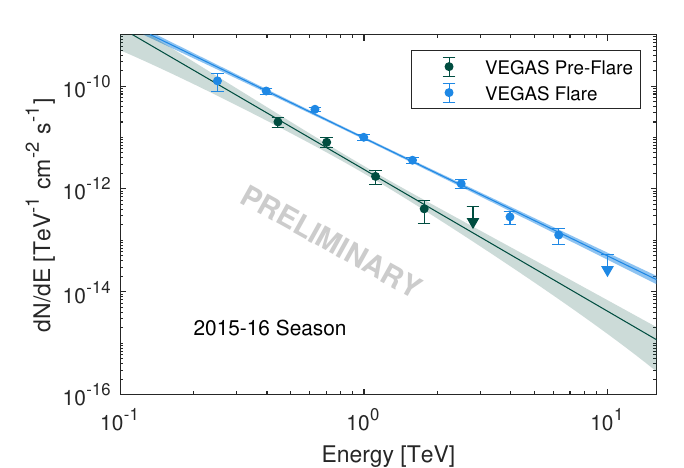}
    \includegraphics[width=77mm,trim={4.5cm 10cm 5.5cm 10cm},clip]{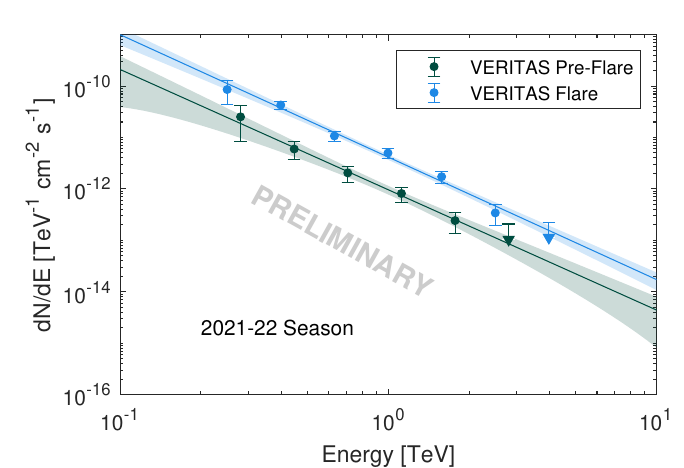}
    \caption{Differential energy spectra of 1ES 2344+514 during pre-flaring and flaring periods for each season.}
    \label{fig:spec}
\end{figure}

Time-averaged spectra are shown in Figure \ref{fig:spec}. Two spectra are shown for each season, split into pre-flare and flare data (57373-57374 MJD and 59546-59551 MJD). They are well modeled by a power-law. Table \ref{tab:spec} lists the fitted parameter values. The analysis was performed in two custom VERITAS packages to confirm results and excellent agreement was found between them.

\begin{figure}
    \centering
    \includegraphics[width=100mm,trim={2.5cm 7.5cm 3.5cm 6cm},clip]{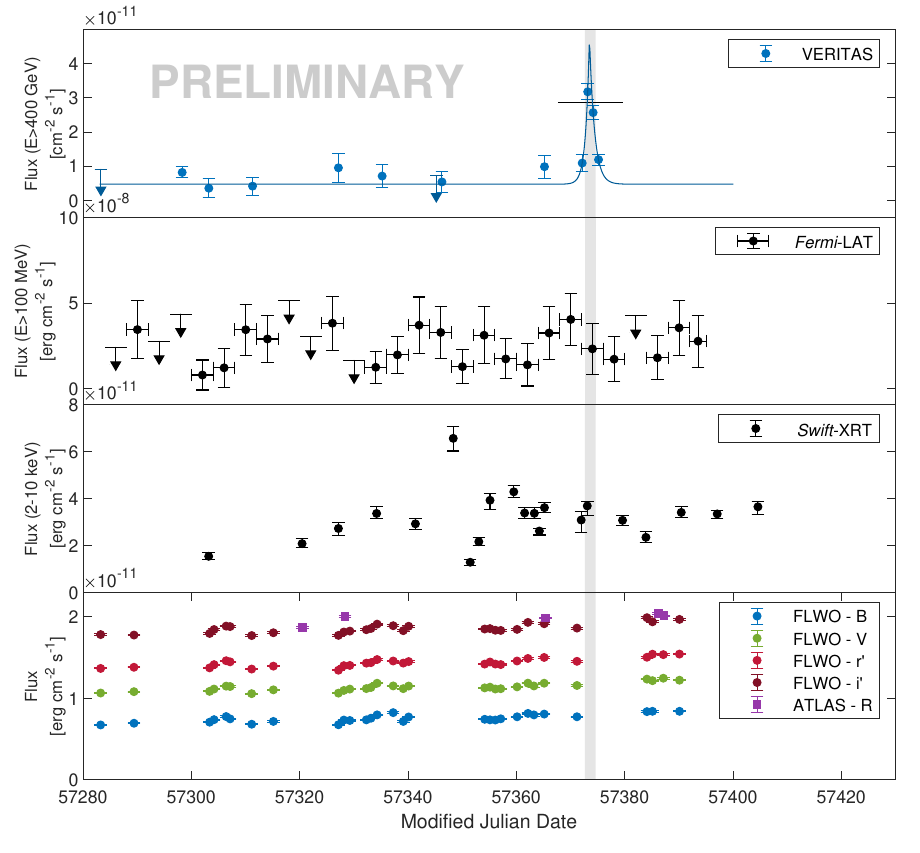}
    \caption{Multiwavelength light curves for 1ES 2344+514 during the 2015-16 season. A Bayesian block analysis of the VERITAS data identifies the flaring region shaded in gray. The average VERITAS flux during the flare is shown as a horizontal line with a value of $(2.9\pm0.3)\times10^{-11}$ cm$^{-2}$ s$^{-1}$.}
    \label{fig:lc15}
\end{figure}

\begin{figure}
    \centering
    \includegraphics[width=100mm,trim={2.5cm 7.5cm 3.5cm 6cm},clip]{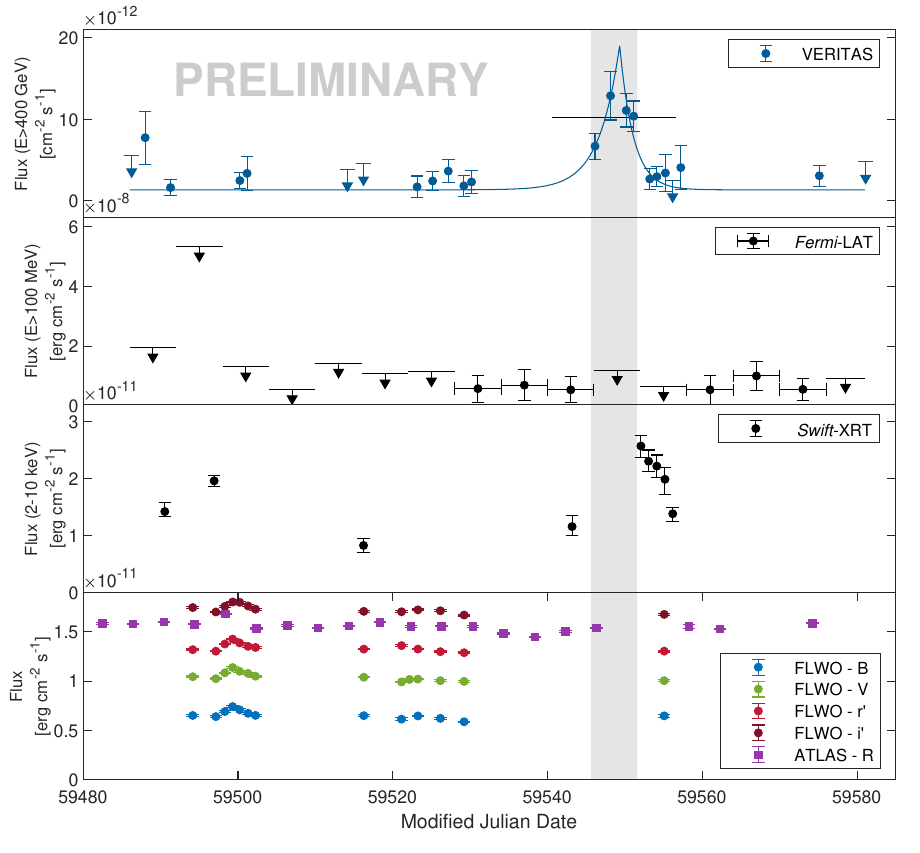}
    \caption{Multiwavelength light curves for 1ES 2344+514 during the 2021-22 season. A Bayesian block analysis of the VERITAS data identifies the flaring region shaded in gray. The average VERITAS flux during the flare is shown as a horizontal line with a value of $(1.03\pm0.13)\times10^{-11}$ cm$^{-2}$ s$^{-1}$.}
    \label{fig:lc21}
\end{figure}

The daily integrated fluxes are shown as light curves for each season in Figure \ref{fig:lc15} and Figure \ref{fig:lc21}. The VERITAS light curves show clear evidence of flaring above the energy threshold of 400 GeV. In each case, the light curves have good quality baselines prior to flares and extended observations during flaring periods. The 2015-16 season has a pre-flare significance of $11.9\sigma$ and a significance $31.3\sigma$ during the flare.  The 2021-22 season has a pre-flare significance of $6.7\sigma$ and a significance $15.9\sigma$ during the flare. 

\begin{table}[b]
    \centering
    \begin{tabular}{|c|c|c|c|c|}
    \hline
        & $F_0$ [TeV$^{-1}$ cm$^{-2}$ s$^{-1}$] & $\Gamma$ & $\tilde{\chi}^2$ & DOF \\ \hline
        2015 Pre-Flare & $(2.4\pm0.4)\times10^{-12}$ & $2.8\pm0.3$ & 0.7 & 2 \\
        2015 Flare & $(9.8\pm0.6)\times10^{-12}$ & $2.29\pm0.06$ & 1.9 & 6 \\\hline
        2021 Pre-Flare & $(9.6\pm1.9)\times10^{-13}$ & $2.3\pm0.3$ & 0.1 & 3 \\
        2021 Flare & $(4.1\pm0.5)\times10^{-12}$ & $2.38\pm0.15$ & 0.8 & 4 \\\hline
    \end{tabular}
    \caption{Fitted parameters of a power-law model of VERITAS differential fluxes for pre-flaring and flaring periods. The power-law model is defined by $F(E)=F_0 \left(\frac{E}{E_0}\right)^{-\Gamma}$ with a normalization energy, $E_0$, fixed at 1 TeV.}
    \label{tab:spec}
\end{table}

\section{Multiwavelength Observation}
\subsection{Fermi-LAT Observations}

NASA's Fermi Large Area Telescope (\emph{Fermi}-LAT) is a space telescope that is sensitive to gamma rays at energies over $\sim$$30$ MeV. Contemporaneous \emph{Fermi}-LAT data (57284-57395 MJD and 59487-59581 MJD) were reduced using \texttt{fermipy} (v1.1.6, and ScienceTools v2.2.0) with the Pass-8 (P8R3\_SOURCE\_V3) instrument response functions \cite{pass8}. The data was filtered as follows. Events from the “Source” class (evclass=128) and from both the front and back (evtype=3) between 100 MeV and 1 TeV within a 15 degree radius of the source location were selected. The data was also then split into PSF event types of which there are 4, representing quartiles that are determined by the quality of the reconstructed photon direction. A zenith cut of $90^{\circ}$ was applied to remove contamination from the Earth’s limb. Another cut was applied to remove times when the Sun was within $5^{\circ}$ of the source. A binned likelihood analysis was performed on the events passing these criteria. The normalization of the sources within $3^{\circ}$ of the source were freed. All parameters of sources with test statistic (TS) above 10 were freed. The isotropic and Galactic diffuse components were freed. Fixed parameters were frozen to their 4FGL-DR3 (gll\_psc\_v28.xml) catalog values \cite{fgl}. Each bin of the light curve has one free parameter — the normalization of the source. The light curves for each season are shown in Figures \ref{fig:lc15} and \ref{fig:lc21}. No evidence of flaring is observed during either VERITAS flare.

\subsection{Swift-XRT Observations}

The Neil Gehrels Swift Observatory (\emph{Swift}) \cite{swift} is a space-based observatory, with an X-ray Telescope (\emph{Swift}-XRT \cite{swift2}) sensitive to X-rays in the 0.3-10 keV range. Contemporaneous observations taken by \emph{Swift}-XRT (57303-57397 MJD and 59490-59556 MJD) were reprocessed using the \texttt{HEAsoft} (v6.31.1) software package. The XRT observations were cleaned and calibrated using \texttt{xrtpipeline} (\cite{swiftste}, v0.13.7).

Observations were taken in the window timing (WT) and photon counting (PC) modes and corrected for pile-up using the methods described in \cite{swift3}. A circular "on" region centered on 1ES 2344+514 with a radius of 20 pixels ($\sim$$47''$), corresponding to 90\% containment of the point spread function of a 1.5 keV photon, was used. For PC mode observations, an annular background region with a $75''$ and $150''$ inner and outer radius, respectively, centered on 1ES 2344+514 was used. For WT mode an annular background with 80 pixel and 100 pixel inner and outer radius was used to ensure a background size of 20 pixel. The BACKSCAL parameter was appropriately adjusted using the \texttt{grppha} command.

Data were grouped using the \texttt{grppha} command such that bad channels were excluded, events outside of the 0.3-10 keV range were excluded and changes were grouped together to ensure a minimum of 20 counts per bin to allow for $\chi^2$-based fitting. Ancillary response files were generated using the \texttt{xrtmkarf} protocol.

Each observation was fit with a deabsorbed power law model (\emph{phabs * power-law}), with the neutral hydrogen column density (N$_{H1}$) frozen to the value measured by the HI 4 Pi survey \cite{swift4} of $1.40\times10^{21} \mathrm{cm}^2$. This fit was performed using the \texttt{PyXSpec} interface to the \texttt{XSpec} package (v12.13.0.c). The integral flux in the 2-10 keV band for each season are shown in Figures \ref{fig:lc15} and \ref{fig:lc21}. Variability is generally observed throughout the light curves. The tail of the flare is observed in Figure \ref{fig:lc21} because the VERITAS veritas flare triggered Swift observations.

\subsection{Optical Observations}

Optical observations were collected by the FLWO 1.2-m telescope \cite{FLWO} and the Asteroid Terrestrial-impact Last Alert System (ATLAS) project \cite{ATLAS1, ATLAS2}. ATLAS is a high-cadence all-sky survey system of four telescopes located in Hawaii, Chile, and South Africa. Contemporanteous FLWO observations (57283-57390 MJD and 59494-59555 MJD) were taken in Harris B, Harris V, SDSS r$^\prime$, SDSS i$^\prime$ filters. The contemporanteous ATLAS observations were taken through an R-filter (57320-57387 MJD and 59482-59574 MJD). In both cases an automated pipeline was run to reduce the data using aperture photometry and nearby stars were used to estimate the systematic uncertainty in the magnitudes. The daily integrated fluxes are shown in Figures \ref{fig:lc15} and \ref{fig:lc21}.

\section{Results and Discussion}

Our light curves for each season are shown in Figures \ref{fig:lc15} and \ref{fig:lc21}. The VERITAS light curves show clear evidence of flaring above the energy threshold of 400 GeV. Interestingly, there is no evidence of simultaneous flaring in the \emph{Fermi}-LAT or optical light energy bands. Because of the VERITAS flare in 2021-22, \emph{Swift}-XRT observations were scheduled, and they show the relaxation of a flare within the energy range 2-10 keV. The \emph{Swift}-XRT observation of the 2015-16 flare is early in the flare and does not show evidence of flaring. We note, however, that \emph{Swift}-XRT generally shows variability throughout both seasons.

For both seasons, the VERITAS light curves were modeled by a time-dependent flux defined by 
\begin{equation}
    F(t)=
    \begin{cases}
        F_{max}e^{\frac{t-t_0}{\tau_1}}+F_{B} & t<t_0\\
        F_{max}e^{-\frac{t-t_0}{\tau_2}}+F_{B} & t>t_0
    \end{cases}
\end{equation}
where $F_{max}$ is the peak flux, $t_0$ identifies the time of the flare, $\tau$ characterizes the duration of heating and cooling periods, and $F_B$ is the baseline non-flaring flux. The fitted parameters are presented in Table \ref{tab:lc}. The average measured flux was 6.1 and 7.6 times higher than that of the fitted baseline flux $F_B$ in 2015-16 and 2021-22, respectively. The baseline flux of 2015-16 is 4 times larger than that of 2021-22, showing significant variability between seasons.

\begin{table}[b]
    \centering
    \begin{tabular}{|c|c|c|c|c|c|c|c|}
        \hline
        & $t_0$ & $F_{max}$ & $\tau_1$ & $\tau_2$ & $F_B$ & $\tilde{\chi}^2$ & DOF \\
        & MJD & $10^{-11}[$cm$^{-2} $s$^{-1}]$ & [day] & [day] & $10^{-12} [$cm$^{-2} $s$^{-1}]$ & & \\\hline
        $2015$ & $57373.45 \pm 0.17$ & $4.3\pm1.0$ & $0.7\pm0.3$ & $0.9\pm0.3$ & $4.69\pm1.17$ & 2.0 & 8 \\
        $2021$ & $59549.3 \pm 0.7$ & $1.8\pm0.6$ & $2.7\pm1.8$ & $1.8\pm0.6$ & $1.4\pm0.4$ & 1.5 & 18 \\\hline
    \end{tabular}
    \caption{Fitted parameters of the light curve model for VERITAS.}
    \label{tab:lc}
\end{table}

\begin{figure}
    \includegraphics[width=77mm,trim={2cm 7cm 2.5cm 8cm},clip]{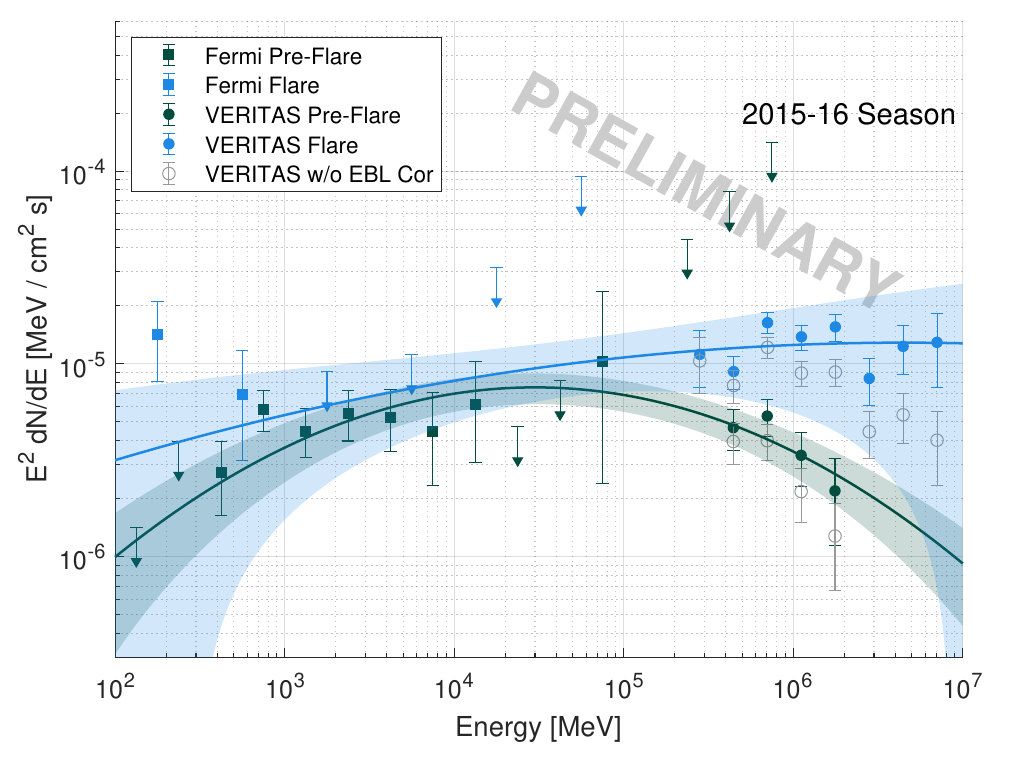}
    \includegraphics[width=77mm,trim={2cm 7cm 2.5cm 8cm},clip]{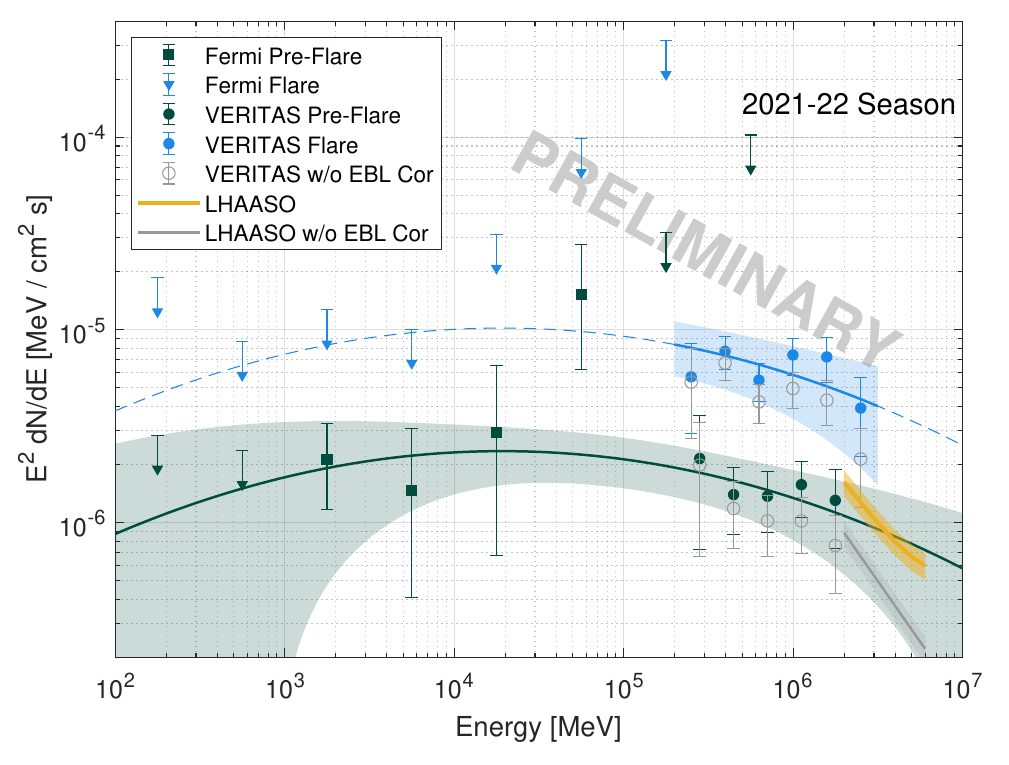}
    \caption{Spectral energy distributions for 1ES 2344+514 during pre-flaring and flaring periods for the (left) 2015-16 season and (right) 2021-22 season.}
    \label{fig:sed}
\end{figure}

Our combined spectral energy distribution (SED) for 2015-16 is shown in Figure \ref{fig:sed} for the pre-flare and flare time periods. All VERITAS data are corrected for attenuation by the extragalactic background light (EBL) \cite{ebl}. For reference, the uncorrected spectral measurements are also included in the plot. Both pre-flare and flare fluxes are fit with a log-parabola model defined by $F(E)=F_0 \left(\frac{E}{E_0}\right)^{-\Gamma-\beta ln(E/E_0)}$, where $F_0$ is the flux at the normalization energy $E_0$, $\Gamma$ is the spectral index, and $\beta$ is the parameter of curvature. The values of these fitted model parameters are given in Table \ref{tab:sed}. Interestingly, the flare and pre-flare butterflies overlap in the \emph{Fermi}-LAT energy range. This implies that there was no flare at these energies, in agreement with the \emph{Fermi}-LAT light curve shown in Figure \ref{fig:lc15}.

Our combined SED for 2021-22 is shown in Figure \ref{fig:sed} for the pre-flare and flare time periods. VERITAS fluxes and the LHAASO \cite{lhaaso} power-law spectrum are corrected for EBL attenuation \cite{ebl}. As before, pre-flare spectral measurements of VERITAS and \emph{Fermi}-LAT were fit to a log-parabola model and the fitted parameters are reported in Table \ref{tab:sed}. The 2021-22 LHASSO measurement is in good agreement with the pre-flare fit in Table \ref{tab:sed}. Due to low statistics at \emph{Fermi}-LAT energies, the pre-flare parameters were re-normalized to the flaring VERITAS fluxes. The $\Tilde{\chi}^2$ shows good agreement with the VERITAS fluxes. \emph{Fermi}-LAT did not see flaring during this period, so the most likely model during the flare should fall between these two fits.

\begin{table}[]
    \centering
    \begin{tabular}{|c|c|c|c|c|c|}
         \hline
         & $F_0$ & $\Gamma$ & $\beta$ & $\tilde{\chi}^2$ & DOF \\
         & MeV$^{-1}$ cm$^{-2}$ s$^{-1}$ & & & & \\\hline
         2015 Pre-Flare & $(3.0\pm0.5)\times10^{-15}$ & $2.06\pm0.04$ & $0.062\pm0.016$ & 0.95 & 11 \\
         2015 Flare & $(4.0\pm1.3)\times10^{-15}$ & $1.89\pm0.13$ & $0.012\pm0.025$ & 2.1 & 7 \\\hline
         2021 Pre-Flare & $(2.1\pm0.6)\times10^{-16}$ & $2.12\pm0.07$ & $0.036\pm0.039$ & 0.4 & 6 \\
         2021 Flare & $(9.3\pm2.9)\times10^{-16}$ &  &  & 0.8 & 5 \\\hline
    \end{tabular}
    \caption{Fitted parameters of log-parabola model to each SED for pre-flaring and flaring periods using \textit{Fermi}-LAT and VERITAS data. Normalization energy $E_0$ is $5 \times10^{4}$ MeV and $10^{5}$ MeV for 2015-16 and 2021-22, respectively.}
    \label{tab:sed}
\end{table}

\section{Summary}

VERITAS observations of 1ES 2344+514 in the 2015-16 and 2021-22 seasons show strong evidence of flaring behavior at very high energies. \emph{Swift}-XRT observations caught the end of the 2021-22 flare. Interestingly, \emph{Fermi}-LAT does not show flaring, potentially limited by short timescale sensitivity. A full study of the synchrotron self-Compton modeling for 1ES 2344+514 will be discussed in a future VERITAS publication.

\section{Acknowledgements}
This research is supported by grants from the U.S. Department of Energy Office of Science, the U.S. National Science Foundation and the Smithsonian Institution, by NSERC in Canada, and by the Helmholtz Association in Germany. This research used resources provided by the Open Science Grid, which is supported by the National Science Foundation and the U.S. Department of Energy's Office of Science, and resources of the National Energy Research Scientific Computing Center (NERSC), a U.S. Department of Energy Office of Science User Facility operated under Contract No. DE-AC02-05CH11231. We acknowledge the excellent work of the technical support staff at the Fred Lawrence Whipple Observatory and at the collaborating institutions in the construction and operation of the instrument.

C. Poggemann is generously supported by the William and Linda Frost Fund through the Bailey College of Science and Mathematics at the California Polytechnic State University, San Luis Obispo.

\clearpage

\section*{Full Author List: VERITAS Collaboration}

\scriptsize
\noindent
A.~Acharyya$^{1}$,
C.~B.~Adams$^{2}$,
A.~Archer$^{3}$,
P.~Bangale$^{4}$,
J.~T.~Bartkoske$^{5}$,
P.~Batista$^{6}$,
W.~Benbow$^{7}$,
J.~L.~Christiansen$^{8}$,
A.~J.~Chromey$^{7}$,
A.~Duerr$^{5}$,
M.~Errando$^{9}$,
Q.~Feng$^{7}$,
G.~M.~Foote$^{4}$,
L.~Fortson$^{10}$,
A.~Furniss$^{11, 12}$,
W.~Hanlon$^{7}$,
O.~Hervet$^{12}$,
C.~E.~Hinrichs$^{7,13}$,
J.~Hoang$^{12}$,
J.~Holder$^{4}$,
Z.~Hughes$^{9}$,
T.~B.~Humensky$^{14,15}$,
W.~Jin$^{1}$,
M.~N.~Johnson$^{12}$,
M.~Kertzman$^{3}$,
M.~Kherlakian$^{6}$,
D.~Kieda$^{5}$,
T.~K.~Kleiner$^{6}$,
N.~Korzoun$^{4}$,
S.~Kumar$^{14}$,
M.~J.~Lang$^{16}$,
M.~Lundy$^{17}$,
G.~Maier$^{6}$,
C.~E~McGrath$^{18}$,
M.~J.~Millard$^{19}$,
C.~L.~Mooney$^{4}$,
P.~Moriarty$^{16}$,
R.~Mukherjee$^{20}$,
S.~O'Brien$^{17,21}$,
R.~A.~Ong$^{22}$,
N.~Park$^{23}$,
C.~Poggemann$^{8}$,
M.~Pohl$^{24,6}$,
E.~Pueschel$^{6}$,
J.~Quinn$^{18}$,
P.~L.~Rabinowitz$^{9}$,
K.~Ragan$^{17}$,
P.~T.~Reynolds$^{25}$,
D.~Ribeiro$^{10}$,
E.~Roache$^{7}$,
J.~L.~Ryan$^{22}$,
I.~Sadeh$^{6}$,
L.~Saha$^{7}$,
M.~Santander$^{1}$,
G.~H.~Sembroski$^{26}$,
R.~Shang$^{20}$,
M.~Splettstoesser$^{12}$,
A.~K.~Talluri$^{10}$,
J.~V.~Tucci$^{27}$,
V.~V.~Vassiliev$^{22}$,
A.~Weinstein$^{28}$,
D.~A.~Williams$^{12}$,
S.~L.~Wong$^{17}$,
and
J.~Woo$^{29}$\\
\\
\noindent
$^{1}${Department of Physics and Astronomy, University of Alabama, Tuscaloosa, AL 35487, USA}

\noindent
$^{2}${Physics Department, Columbia University, New York, NY 10027, USA}

\noindent
$^{3}${Department of Physics and Astronomy, DePauw University, Greencastle, IN 46135-0037, USA}

\noindent
$^{4}${Department of Physics and Astronomy and the Bartol Research Institute, University of Delaware, Newark, DE 19716, USA}

\noindent
$^{5}${Department of Physics and Astronomy, University of Utah, Salt Lake City, UT 84112, USA}

\noindent
$^{6}${DESY, Platanenallee 6, 15738 Zeuthen, Germany}

\noindent
$^{7}${Center for Astrophysics $|$ Harvard \& Smithsonian, Cambridge, MA 02138, USA}

\noindent
$^{8}${Physics Department, California Polytechnic State University, San Luis Obispo, CA 94307, USA}

\noindent
$^{9}${Department of Physics, Washington University, St. Louis, MO 63130, USA}

\noindent
$^{10}${School of Physics and Astronomy, University of Minnesota, Minneapolis, MN 55455, USA}

\noindent
$^{11}${Department of Physics, California State University - East Bay, Hayward, CA 94542, USA}

\noindent
$^{12}${Santa Cruz Institute for Particle Physics and Department of Physics, University of California, Santa Cruz, CA 95064, USA}

\noindent
$^{13}${Department of Physics and Astronomy, Dartmouth College, 6127 Wilder Laboratory, Hanover, NH 03755 USA}

\noindent
$^{14}${Department of Physics, University of Maryland, College Park, MD, USA }

\noindent
$^{15}${NASA GSFC, Greenbelt, MD 20771, USA}

\noindent
$^{16}${School of Natural Sciences, University of Galway, University Road, Galway, H91 TK33, Ireland}

\noindent
$^{17}${Physics Department, McGill University, Montreal, QC H3A 2T8, Canada}

\noindent
$^{18}${School of Physics, University College Dublin, Belfield, Dublin 4, Ireland}

\noindent
$^{19}${Department of Physics and Astronomy, University of Iowa, Van Allen Hall, Iowa City, IA 52242, USA}

\noindent
$^{20}${Department of Physics and Astronomy, Barnard College, Columbia University, NY 10027, USA}

\noindent
$^{21}${ Arthur B. McDonald Canadian Astroparticle Physics Research Institute, 64 Bader Lane, Queen's University, Kingston, ON Canada, K7L 3N6}

\noindent
$^{22}${Department of Physics and Astronomy, University of California, Los Angeles, CA 90095, USA}

\noindent
$^{23}${Department of Physics, Engineering Physics and Astronomy, Queen's University, Kingston, ON K7L 3N6, Canada}

\noindent
$^{24}${Institute of Physics and Astronomy, University of Potsdam, 14476 Potsdam-Golm, Germany}

\noindent
$^{25}${Department of Physical Sciences, Munster Technological University, Bishopstown, Cork, T12 P928, Ireland}

\noindent
$^{26}${Department of Physics and Astronomy, Purdue University, West Lafayette, IN 47907, USA}

\noindent
$^{27}${Department of Physics, Indiana University-Purdue University Indianapolis, Indianapolis, IN 46202, USA}

\noindent
$^{28}${Department of Physics and Astronomy, Iowa State University, Ames, IA 50011, USA}

\noindent
$^{29}${Columbia Astrophysics Laboratory, Columbia University, New York, NY 10027, USA}

\end{document}